\documentclass[twocolumn]{revtex4}

\usepackage{amsmath, amsfonts, amssymb, graphicx,color}
\usepackage[T1]{fontenc} 
\usepackage[dvipsnames]{xcolor}

\newcommand{\equal}{Corresponding authors. These authors contributed equally.}
\newcommand{\ENS}{{Laboratoire de physique de l'\'Ecole normale sup\'erieure,
		CNRS, PSL University, Sorbonne Universit\'e, and Universit\'e 
		Paris-Cit\'e, 75005 Paris, France}}

\newcommand{\TM}{}

\newcommand{\ha}[1]{{#1}}
\newcommand{\add}[1]{{#1}}
\newcommand{\PE}{\mathcal{P}(E)}
\newcommand{\<}{\langle}
\renewcommand{\>}{\rangle}

\begin{document}
\affiliation{\ENS}
\author{Henry Alston}
\affiliation{\ENS}
\author{Thierry Mora}
\thanks{\equal}
\affiliation{\ENS}
\author{Aleksandra M. Walczak}
\thanks{\equal}
\affiliation{\ENS}

\newcommand{\deftitle}{{Neutralization titers reveal the structure of polyclonal antibody responses}}

\title{\deftitle}

\begin{abstract}
The composition of a polyclonal antibody response is hard to measure experimentally but contains vital information about the robustness of immunity. Here, we argue that the statistics of neutralization titers alone can be used to make quantitative predictions about the composition of the response, circumventing challenges arising through sequencing and monoclonal antibody expression. We show that the response against influenza within a cohort can be either driven by a collective phenomenon where many antibodies contribute to neutralization, or dominated by just a few strong binders, leading to a broad distribution of titers across individuals described by a Gumbel distribution from extreme value theory. Comparing titers across cohorts, we find that Gumbel statistics {accurately describe} individuals prior to an immune challenge. We propose an equilibrium binding model that quantitatively captures titer data and illustrates the structure of the polyclonal response. Our approach extends generically to immune responses to other pathogens.

\end{abstract}

\maketitle

\section{Introduction}

Antibody responses are central to the adaptive immune system, enabling targeting of specific antigens through acquired immunological memory upon repeated pathogen exposure. These responses are inherently polyclonal, composed of multiple distinct clonotypes generated through V(D)J recombination, somatic hypermutation, and clonal selection, ensuring broad and specific pathogen recognition \cite{Jerne1955, Burnet1957, Tonegawa1983, Alt1992}.

But how many antibodies define a polyclonal response? Neutralization titers describe the ability of polyclonal sera to inhibit viral infection. However, a single titer measurement does not reflect the underlying clonal composition of the antibody response. Whether neutralization reflects a broad collective effort or is dominated by a few clonotypes is essential for assessing immune robustness \cite{Lee2019b, Schnaack2021, Greaney2021b, MunozAlia2021, Chardes2022}. Narrow responses, even when potent, are potentially susceptible to viral escape through single-point mutations \cite{Starr2020, Greaney2021a, Cao2022, Cao2022b}, in principle limiting the durability and breadth of protection. In contrast, broadly polyclonal responses provide more resilient immunity, as viral escape may require coordinated mutations across multiple epitopes \cite{Scheid2009, Yu2022}. 

Experimentally, several approaches have been developed to dissect the polyclonal serum response, combining sequencing of B-cell repertoires with proteomic identification of serum antibodies and subsequent monoclonal expression to characterize individual clonotypes \cite{Wine2013, Lee2016}. Yet this workflow faces three fundamental challenges. First, incomplete and biased sampling limits how well the cellular repertoire reflects the antibodies actually circulating in serum, particularly because key populations such as bone-marrow plasma cells are often missed, and because the native heavy-light chain pairing cannot always be reliably recovered. Second, imperfect molecular identification hampers proteomic resolution, as bottom-up mass spectrometry often fails to detect the unique CDRH3 peptides required to distinguish clonotypes, and antigen-affinity enrichment and pepsin digestion introduce additional noise that can obscure true binders. Third, functional reconstruction is incomplete since monoclonal antibodies may not recapitulate the cooperative and competitive interactions that shape neutralization in the native serum environment. These challenges make it difficult to obtain a complete and accurate picture of which clonotypes mediate serum activity, {limiting these bottom-up approaches to a few studies} \cite{Wine2013, Lee2016}.

While the fine structure of serum antibody repertoires remains difficult to resolve, neutralization assays directly measure the ability of polyclonal immune responses to neutralize their target and thus provide the gold-standard functional readout in immunology. Recent advances now allow these titers to be measured at high throughput against many viral variants \cite{Kikawa2025, Kikawa2025b}, but still avoid the complexity of repertoire-level sequencing.

In this work, we demonstrate that variations in neutralization titers across a cohort can be used to infer key features of the immune response to viral infection. We construct an equilibrium model for antibody-virus binding that connects neutralization titers to the constituent binding affinities in sera. Through computational analysis, we show how this approach reveals the underlying structure of the response. The framework is agnostic to viral identity, providing a general method for analyzing polyclonal responses to infection.

\section{Results}

\subsection{Children and post-vaccination adults exhibit different distributions for measured neutralization titers}

\begin{figure*}[t]
    \includegraphics[width=\linewidth]{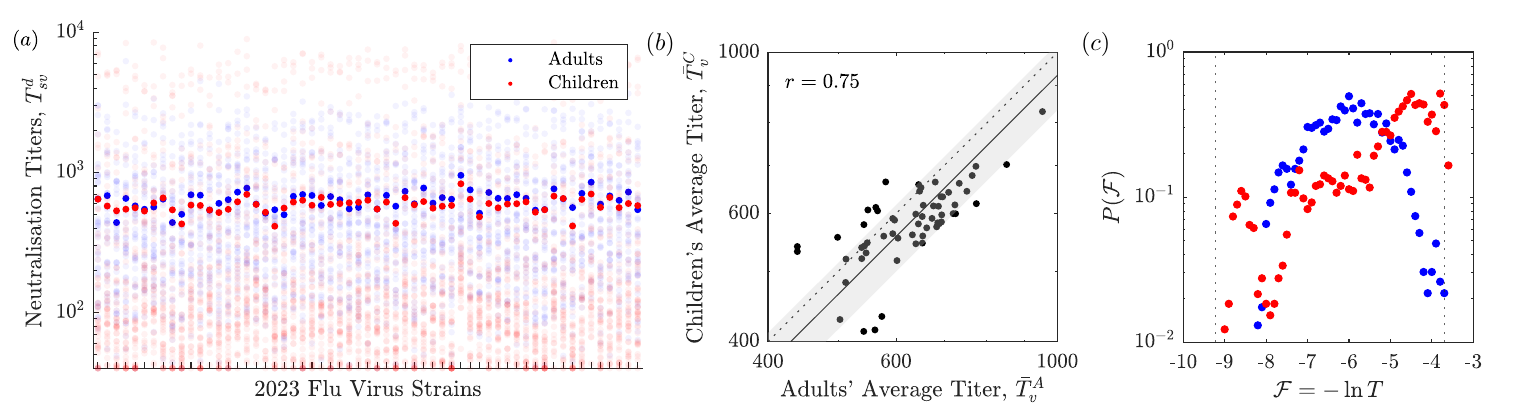}
    \caption{\textbf{Different distributions for titers across cohorts ---} (a) Neutralization titers $T_{sv}^d$ for serum $s$ against virus $v$ for two datasets $d=\{A,C\}$ against influenza H3N2 strains from 2023. (C)hildren are plotted in red, post-vaccination (A)dults in blue. Light dots show individual sera and dark dots the average over all sera samples for a specific viral strain. (b) The two datasets have comparable mean titers across virus strains: average ratio $\bar{T}_v^C/\bar{T}_v^A$ of 0.93 (solid line) with standard deviation $0.11$ (grey area) and strong positive correlation ($r=0.75$). Dashed line is $\bar{T}_v^A=\bar{T}_v^C$ and is within the grey area. (c) The two datasets exhibit different distributions of $\mathcal{F}=-\ln T$ where $T\in\{T_{sv}^d\}$ as defined in Eq.~\eqref{eq:rescaleT}: adults (blue) exhibit Gaussian statistics, whereas the children (red) appear non-Gaussian. }
    \label{fig:data}
\end{figure*}

We first analyze data from Kikawa et al.~\cite{Kikawa2025} which quantitatively measures serum antibody titers against 78 influenza H3N2 strains through a high-throughput sequencing-based neutralization assay. Briefly, the method utilizes a barcoded viral library, where each hemagglutinin (HA) variant is tagged with a unique nucleotide sequence. These barcoded viruses are pooled and incubated with serially diluted human sera from 40-fold to 10000-fold dilution. Following infection of cells,
viral RNA is extracted and sequenced, and the relative abundance of each variant estimated by counting barcodes.
The neutralization titer $T$ for a serum-virus pair is defined as the dilution at which the serum retains 50\% of its undiluted neutralization efficacy. In the experiments of Kikawa et al.~\cite{Kikawa2025}, the undiluted serum concentration is $c_0=667$\,nM. The concentration corresponding to half-maximal efficacy is then denoted $c_{50} = c_0 / T$, where $T$ is the measured titer. Higher titers $T$ denote more potent sera. We denote by $T_{sv}^d$ the measured titer for serum $s$ against virus strain $v$ for dataset $d=\{A, C\}$ where $d$ denotes either {(i) post-vaccination (A)dults or (ii) (C)hildren.} \add{The number of adult sera is 39, compared to 56 children and the vaccination history of the children is not known \cite{Kikawa2025}.}

In Fig.\,\ref{fig:data}(a), we plot the raw data for titers measured for a broad range of sera against flu virus strains from 2023 \cite{Kikawa2025}. The results for sera taken from post-vaccination adults are plotted in blue, and from children in red. Individual sera are plotted in light colors, whereas the dark colored point is the average of the titers for virus $v$, $\bar{T}_v^A$ for adults, and $\bar{T}_v^C$ for children.
By eye, the average titers $\bar{T}_v^A$ and $\bar{T}_v^C$ are comparable across viruses. We quantify this in Fig.\,\ref{fig:data}(b) by plotting the mean titers of the adults ($x$-axis) against those of the children ($y$-axis) for each virus, revealing a strong correlation between the two (Pearson $r=0.75$).
The log ratio $\mathcal{R}=\ln\left(\bar{T}_v^C/\bar{T}_v^A\right)$ between children and adults
has mean $-0.07$ and standard deviation $0.11$:
adult sera perform better in neutralizing the influenza strains considered, but the child sera perform almost as well on average and even outperforms the adult sera in some cases.

Beyond comparing averages between the two datasets, the large number of measured titers enables a more quantitative analysis of the distribution of neutralization abilities across cohorts. Titers within dataset $d$ are not immediately comparable due to the inherent variability between viruses. We thus re-scale the titers within each virus to offset this variability
by replacing  ${T}_{sv}^d$ by
\begin{equation}\label{eq:rescaleT}
   T_{sv}^d  \leftarrow T_{sv}^d \times \frac{\bar{T}^d}{\bar{T}_v^d},
\end{equation}
where $\bar{T}_v^d$ is the average titer measured against virus $v$ in dataset $d$ and $\bar{T}^d$ the average titer in dataset $d$.
This ensures
that the average titer against each virus strain is the same for all strains. By offsetting the variability between viruses, we treat our data set as if they were titers measured for multiple independent sera against the same virus. 

Finally, we define $\mathcal{F}_{sv}^d=-\ln{T}_{sv}^d$ and
build a distribution for $\mathcal{F}$ over $s$ and $v$ for each dataset $d$,
for the Adult and Children datasets (Fig.\,\ref{fig:data}(c)). \add{While different influenza strains may be neutralized by distinct subsets of antibody clonotypes, the distributions of neutralization titers are nevertheless remarkably consistent across the strains examined (Fig.\,\ref{fig:data}(a)). We therefore adopt a coarse-grained statistical perspective, in which responses across strains are treated as samples from the same underlying distribution. Under this assumption, concatenating data across strains provides a practical means of characterizing the statistics of inter-individual variability without requiring the same clonotypes to mediate neutralization of every strain.}

The $\mathcal{F}$-distributions vary significantly between post-vaccinated adults and children, both in their mean ($-6.0$ vs $-5.0$) and standard deviations ($0.9$ vs $1.0$).
They also differ in their shape: while they appear to be Gaussian in post-vaccinated adults, this is clearly not the case for the children, where a large asymmetry is observed in the distribution. We quantify this through D'Agostino's test, a statistical test for zero skewness: the adult data exhibits a non-significant skew of $-0.06$ ($p=0.23$). The child data has a skewness of $-0.80$: a significant departure from Gaussian statistics ($p\approx10^{-70}$).

Intuitively, one might expect the titers to be determined through some average of the neutralization ability of the constituent antibodies. The central limit theorem would suggest that fluctuations in such averages generically appear as Gaussian when the average is taken over many quantities (i.e. many antibody clonotypes in the sera). While this picture seems to capture the fluctuations in post-vaccinated adults, the child sera indicate a clear departure from the central limit theorem. We seek to understand what causes this departure and what it tells us about the underlying structure of the polyclonal response. 

\subsection{Equilibrium binding model connects serum structure to neutralization ability}

\begin{figure*}
    \centering
    \includegraphics[width=\linewidth]{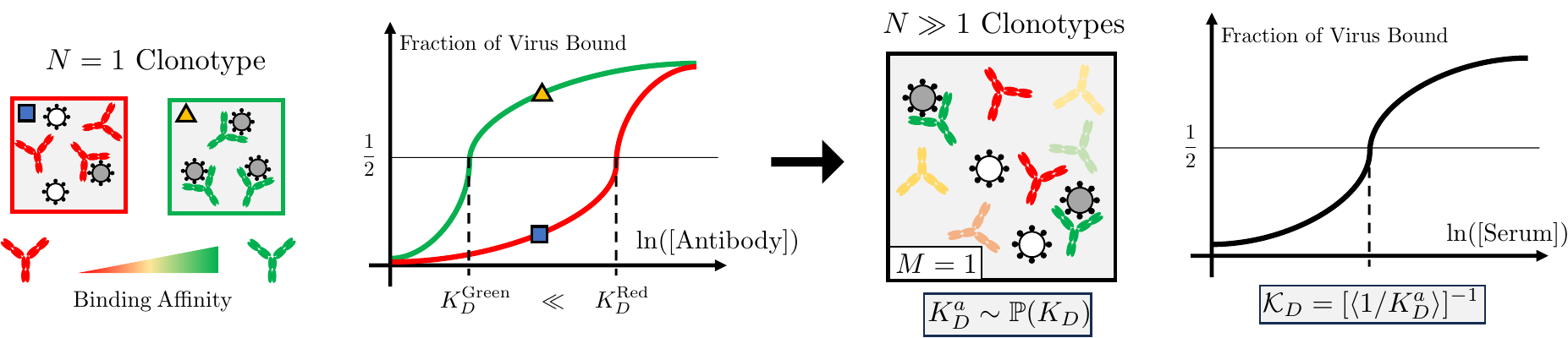}
    \caption{\textbf{Schematic of the equilibrium binding model {\TM for a single epitope} ---} $N$ antibody clonotypes each bind to one epitope of the virus with dissociation constants $\{K^a_D\}$. For antibody clonotypes, $K_D^a$ denotes the concentration when 50\% of the virus is bound. We can define an effective dissociation constant for the serum $\mathcal{K}_D$ (i.e.~a mixture of $N$ clonotypes) as the harmonic mean of the individual $K_D^a$. If all clonotypes bind to one epitope on the virus, $\mathcal{K}_D$ is directly linked to the neutralization titer through {\TM $c_{50}=c_0/T=\Lambda \mathcal{K}_D$, with $\Lambda$ a constant.} }
    \label{fig:model}
\end{figure*}

We begin with an equilibrium antibody-virus binding model. The aim is to develop a physical, microscopic interpretation for the quantity $\mathcal{F}=-\ln{T}$ and thus to understand what the distributions in Fig.\,\ref{fig:data}(c) teach us about the polyclonal response.

We model the serum response as arising from $N$ antibody clonotypes, each descending from a single B-cell lineage and treated as a monoclonal antibody with the same epitope specificity and affinity. \add{This amounts to assuming that the variance in affinity within a lineage is much smaller than the variance across lineages, which is shaped by the full history of selection: members of the same lineage inherit a common receptor framework and differ only by a limited number of recent mutations, whereas different lineages originate from distinct founder clones.} We denote by $c$ the total concentration of the serum and by $f_a$ the fraction of the serum made up of clonotype $a$. Thus, clonotype $a$ is at concentration $f_ac$. 

\add{We say that each viral particle is made up of $Q$ identical tiles, each with $M$ distinct epitopes.} Following previous work \cite{Einav2020}, we make the assumption that each antibody clonotype binds to exactly 1 of these $M$ epitopes, neglecting the possibility of binding to a non-specific epitope. We label the specific epitope of clonotype $a$ as $e_a$ and denote by $K_D^a$ the dissociation constant for the antibody and this epitope.

\ha{We first consider the binding probability for a single tile, denoting by $p_{\rm free}(c)$ the probability that it remains unbound when the concentration of the serum is $c$.} 
In Appendix \ref{sec:SIderiv}, we derive an expression for the fraction of free viral particles (along the same lines as Einav \& Bloom~\cite{Einav2020}):
\begin{align}\label{eq:pfreeM}
p_{\rm free} (c)&= \prod_{e=1}^M\left(\cfrac{1}{1 + \sum\limits_{{a:e_a=e}}\frac{f_a c}{K^a_D}}\right)= \prod_{e=1}^M\left(\frac{1}{1 + \frac{ F_e c }{ \mathcal{K}_D^{e}}}\right),
\end{align}
where we have defined the fraction of clonotypes specific to epitope $e$ as $F_e=\sum\limits_{{a:e_a=e}}f_a\leq1$, and $\mathcal{K}_D^e$ is the harmonic mean of dissociation constants binding to epitope $e$: $\mathcal{K}_D^e = \left[F_e^{-1} \sum_{a:e_a=e} f_a/K_D^a \right]^{-1}$. 

For $M=1$, the result simplifies to $p_{\rm free} (c)=[1+c/\mathcal{K}_D]^{-1}$ where $\mathcal{K}_D$ is the harmonic mean of all dissociation constants. {For $M>1$, the product in Eq.\,\eqref{eq:pfreeM} results in a fraction with a denominator that is an $M$-order polynomial in $c$. \add{We show in the Supplementary Information that the expression simplifies in extreme cases. Briefly, if the antibodies heavily target a single epitope, then the result for $M=1$ holds qualitatively (as the sum for $\mathcal{K}_D$ will be dominated by the contribution of said epitope). In the opposite limit, where there are many epitopes $M$ each equally targeted (so $F_e\approx 1/M$), then we instead derive the expression $p_{\rm free} (c)\approx e^{-c/\mathcal{K}_D}$.}

\add{In these two limits, we see an exact linear relation between the concentration at which a certain fraction (e.g. 50\%) of viral tiles are unbound and the dissociation constant $\mathcal{K}_D$. We verify numerically that a linear scaling is also observed beyond these limit cases: for many epitopes, broad affinity variability and inhomogeneous serum fractions (see results detailed in the Supplementary Information).} 

\add{We now connect binding at viral tiles to the neutralization of a virus particle. Neutralization often requires cooperative interactions between antibodies across epitopes. We model these interactions phenomenologically through a spin model (each tile is associated a spin {\TM equal to +1 if the tile is bound, -1 if it's not}) such that the re-scaled concentration sets an external field \ha{$h (c/\mathcal{K}_D)\equiv k_BT\ln [(1-p_{\rm free})/p_{\rm free}]/2$} on {\TM each spin, with an additional generic interaction potential between spins to capture cooperativity between antibodies across tiles. That interaction term is concentration-independent and we assume it to be agnostic of clonotype identity.} We can then define the criterion for the viral particle to be neutralized as some complicated {\TM probilitistic} function of the collective spins (which we leave to be generic).}

\add{Ultimately, the resulting probability of neutralization will be a non-linear function {\TM $f$} of $c/\mathcal{K}_D$ {\TM(see Supplemenantary Material)}.
  From our analysis above, it follows that {\TM $c_{50}=c_0/T=\mathcal{K}_Df^{-1}(1/2)$} is proportional to the harmonic mean of the dissociation constants $\mathcal{K}_D$ with a coefficient {\TM$\Lambda\equiv f^{-1}(1/2)$} set by the details of neutralization. Explicitly, we can write
    \begin{equation}
{\TM
  \label{eq:FT}\mathcal{F}=-\ln T = \ln (\mathcal{K}_D/c_0) + \ln\Lambda,
}
    \end{equation}
which crucially signifies that fluctuations in the observed titers across patients $\mathcal{F}$ are mirrored in fluctuations in $\ln\mathcal{K}_D/c_0$. }

We now use Eq.\eqref{eq:FT} to relate $\mathcal{F}=-\ln  T$ to the binding energies for the constituent antibodies (also rescaled by $RT$): \add{$\phi_a \equiv \ln(K^a_D\Lambda/c_0)$.} Using the definition of $\mathcal{K}_D$, we get
\begin{equation}\label{eq:fzm1}
	  T =  \sum_{a=1}^Nf_a \exp(-\phi_a)=  \sum_{a=1}^N \exp(-\phi_a+\ln f_a).
\end{equation}
The titer takes the form of a partition function for a system in the canonical ensemble with discrete energy levels $\phi_a-\ln f_a$, and $\mathcal{F}$ may be interpreted as a free energy \add{(where the energy scale is such that $\mathcal{F}$ is zero for unit titer $T$)}. This exact result explicitly relates the composition of the serum to the neutralization titer. The fluctuations of $\mathcal{F}$ across sera can then be analyzed through calculating statistics for $\mathcal{F}$, a problem with a long history in the study of disordered systems. 

\subsection{Asymmetric titer distributions emerge at high clonotype variability}\label{asymmetric}

In Eq.\,\eqref{eq:fzm1}, we have established a link between the structure of a polyclonal serum (characterized by its clonotype's affinities $\phi_a$ and compositions $f_a$) and its neutralization titer through $\mathcal{F}=-\ln {T}$. We now show how particular compositional structures of the antibody repertoire lead to distinct distribution of titers, and in particular asymmetric ones as observed in the data (Fig.\,\ref{fig:data}(c)).

Each serum within a cohort is characterized by the $N$ energies $E_a=\phi_a-\ln f_a$ in our binding model. \add{To simplify our analysis, we say that each $E_a$ is independently drawn from a Gaussian distribution $\PE$ which is fixed for each serum. 
  This is an oversimplification on two counts. First, the affinities $\phi_a$ are the outcome of affinity maturation, not a random process, as selection generates correlations between clones rather than independent draws.
  Second, the Gaussian form of $\PE$ is not meant to capture the selective process that shapes the affinity distribution, but is instead a minimal phenomenological choice, fixed by only its mean and variance, that lets us ask whether the emergence of non-trivial titer statistics depends on the tail behavior of $\PE$. {\TM We expect the following analyses to hold for other distributions with similar tail behaviour, namely faster than exponential decay.}}

We numerically generate $N_s=10^5$ polyclonal sera, each composed of $N=1,000$ antibodies with $E_a$ against a single virus strain drawn from $\PE$.
To model variability between sera, we assume that each serum is described by a Gaussian distribution $\PE$ of antibodies with a serum-independent variance $\sigma^2$ but a serum-dependent mean $\mu_s$, which is itself normally distributed with mean $\mu$ and variance $\sigma_1^2$ across individuals. \add{This choice to introduce person-to-person variability only in the mean affinity is made to isolate whether the observed asymmetry in titers can be explained by cohort-specific antibody variability alone.}
For each serum, we then evaluate $\mathcal{F}=-\ln T$ through Eq.\,\eqref{eq:fzm1}.

We use these numerical realizations to show how the parameters $\mu$ and $\sigma$ control the statistics of $\mathcal{F}$, with fixed $\sigma_1=0.9$. The mean of $\mathcal{F}$ is plotted in Fig.~\ref{fig:nums}(a) and its skewness, $\<(\mathcal{F}-\<\mathcal{F}\>)^3\>/\<(\mathcal{F}-\<\mathcal{F}\>)^2\>^{3/2}$, in Fig.~\ref{fig:nums}(b). We observe that we can improve the mean effective binding efficacy of the serum (lower $\mathcal{F}$) through either decreasing $\mu$ or increasing $\sigma$. \add{The shift arising due to increasing $\sigma$ is due the creation of more low-energy values $E_a$ that dominate the partition function in Eq.\,\eqref{eq:fzm1}, reducing the average free energy $\mathcal{F}$.}

However, low $\sigma$ generates symmetric distributions for $\mathcal{F}$, with negligible skewness. At larger $\sigma$, we observe asymmetric distributions for $\mathcal{F}$, characterized by a skewness that is significantly less than zero. {\TM We will analyze this behaviour in detail in the next section.} Our model can therefore produce titer distributions that are symmetric (as in the case of the post-vaccinated adults) or asymmetric (as for the children) controlled by $\sigma$, even though $\PE$ is Gaussian and always symmetric. 

\begin{figure}
    \centering
    \includegraphics[width=\linewidth]{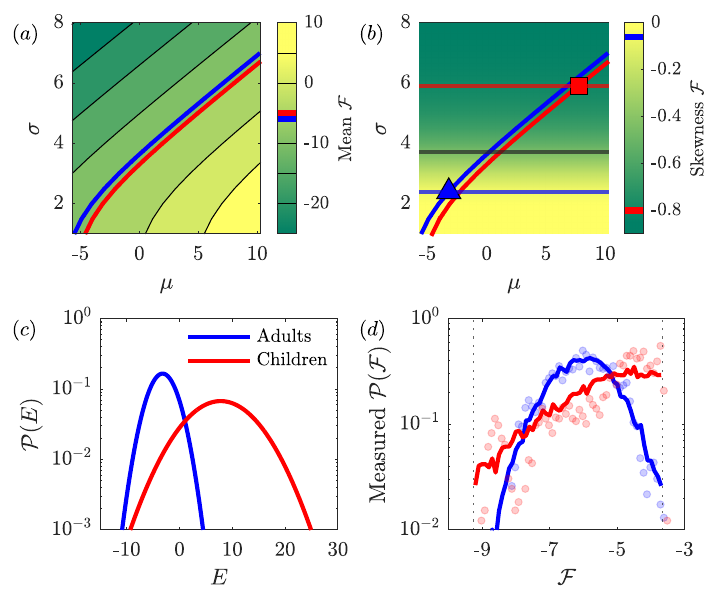}
    \caption{\textbf{Numerical analysis and extracted  distributions of antibody affinities and serum composition $\PE$ ---} We vary the mean $\mu$ and standard deviation $\sigma$ for a Gaussian $\PE$ to see how it sets (a) the mean and (b) the skewness of the resulting distribution of $\mathcal{F}$ across samples. We set $\sigma_1=0.9$.  The black line in (b) corresponds to the transition between the regime when neutralization is dominated by a few and many antibody types. (c) We then fit our model to data.  Matching the mean and skewness of $\mathcal{F}$ for each dataset, we extract a $\mu$ and $\sigma$ defining a Gaussian $\PE$ that fits the adult (blue) and child (red) data. (d) Numerically generated distributions $\mathcal{P}(\mathcal{F})$ (solid lines) for these $\PE$ distributions show good quantitative agreement with experimental data from Fig.~\ref{fig:data}(c) (points). 
}
    \label{fig:nums}
\end{figure}

We now return to the experimental data to confirm if the different titer distributions between post-vaccination adults and children can be explained with our binding model. 
The variance of $\mathcal{F}$ has two sources: the variability in the mean antibody affinity $\mu_s$ across sera, and fluctuations $\sigma$ of the affinity across antibodies within each serum. Since $\mathcal{F}$ results from a sum over $N=1,000$ antibodies, the latter are expected to be small thanks to the law of large numbers, unless $\sigma$ is so large that we enter the regime of extreme values.
We checked that in the parameter regime relevant to the vaccinated adult cohort, where the distribution of $\mathcal{F}$ is symmetric in accordance with the law of large numbers (small $\sigma$), the standard deviation of the numerically generated $\mathcal{F}$ is indeed approximately given by $\sigma_1$ (see Appendix \ref{SIsec:sigma1}). We thus set $\sigma_1=\sqrt{\mathrm{Var}(\mathcal{F})}=0.9$ in vaccinated adults.
We use the same value for $\sigma_1$ for all data sets, such that it sets the inherent variability in $\mu_s$ between any two sera within the same cohort. \add{ It would be interesting to measure this quantity (and also $N$) directly from antibody-level binding data for each cohort. In the absence of such data, we chose to keep these quantities fixed across cohorts, enabling direct comparison of the extracted $\PE$ distributions while avoiding the introduction of additional cohort-specific parameters.}

\add{We then extract the two remaining parameters, $\mu$ and $\sigma$, for each dataset using the results of Fig.~\ref{fig:nums}(a) and (b). In each case, we impose contour lines drawn in blue (post-vaccination adults) and red (children) for the corresponding values of the mean and skewness of $\mathcal{F}$ as measured from the distributions in Fig.~\ref{fig:data}(c). These contours intersect uniquely in each case defining a unique pair ($\mu,\sigma$).} For the post-vaccinated adults, we infer $\mu=-3.2$ and $\sigma=2.4$, and for the children, $\mu=7.8$ and $\sigma =5.9$. The corresponding Gaussian distributions $\PE$ are plotted in Fig.~\ref{fig:nums}(c). \add{This difference in $\sigma$ may reflect the fact that either affinities or the serum fractions are more variable in the child sera than in those of post-vaccinated adults: antibody-level binding data could be useful to distinguish which of the two is important here.} \add{Our choice of $N=1,000$ defines the number of relevant clonotypes constituting the immune response and is consistent with molecular-level studies (see Section H). These parameters (extracted in figure 3) are sensitive to our choice of $N$.}

In Fig.~\ref{fig:nums}(d), we compare the numerically generated distribution of $\mathcal{F}$ for each sets of parameters to the experimentally-measured $\mathcal{F}$ from Fig.~\ref{fig:data}(c), finding good agreement between the two. Note that when comparing to experiments, we discard any numerical data that appears outside of the experimentally measurable range (titers were measured between 40- and 10000-fold dilution), imposing a finite range for $\mathcal{F}$ in Fig.~\ref{fig:nums}(d). Our binding model successfully describes the distributions of titers of both vaccinated adult and children. For adults, a relatively low $\sigma$ results in a symmetric distribution (in blue), while for children a high $\sigma$ results in an asymmetric, negatively skewed distribution (in red).

\subsection{Mapping to Random Energy Model predicts two regimes for sera neutralization}\label{sec:REM}

Our numerical analysis above predicts that asymmetric titer distributions arise when the \add{variability in $E$ is large}. We now seek to gain insight into this phenomenon with a formal analytical treatment of the statistics of $\mathcal{F}$ as arising through Eq.\,\eqref{eq:fzm1}.
This calculation closely follows the approach developed for the Random Energy Model (REM), a paradigmatic model in the study of disordered systems \cite{Derrida1980, Derrida1981}, which has also been applied to protein folding \cite{Bryngelson1987, Frauenfelder1991, Bryngelson1995, Onuchic1997} {and transcription factor binding to DNA \cite{Gerland2002, Aurell2007, Mustonen2008}}. {A similar approach was also recently used to characterize the initial clonal expansion of naive B cells in response to infection \cite{MoranTovar2024}.} The main feature of the REM is the existence of a condensation transition as parameters (such as temperature) are varied, whereby the system ``freezes'' into a few configurations with low energy.

We summarize here the schematic argument of \cite{Derrida1980} adapted to our context. While not all assumptions necessary to the original derivation of the REM results are satisfied in our case, it is still a useful analysis to interpret the behavior of the model.
We define a ``density of states'' for the set of $N$ energies $\{E_a\}$ as $\rho(E)\delta E = \#(\{a\::\:E_a\in[E, E+\delta E)\})$.
On average this density is given by:
\begin{equation}\label{eq:rhoE}
    \langle \rho(E)\rangle = N \PE.
  \end{equation}
For a given set of $N$ antibodies, there exists a minimal and maximal value of $E_a$:
    $E_{\rm min}$ and $E_{\rm max}$.
    This defines a range $(E_{\rm min}, E_{\rm max})$ of represented energies where $\langle \delta E\rho(E)\rangle\geq 1$. When $N$ is large and $\PE$ falls off quickly at the two ends of the distribution, we can assume that $\delta E\rho(E)$ is typically large inside that range, and 0 outside. Since fluctuations of $\delta E\rho(E)$ are of order $\sqrt{\langle \delta E\rho(E)\rangle}$, we can approximate $\delta E\rho(E)\approx \langle \delta E\rho(E)\rangle$ inside the range. With these assumptions Eq.\,\eqref{eq:fzm1} can be written as an integral:
\begin{align}\label{eq:saddle}
     T&\approx {\int_{E_{\rm min}}^{E_{\rm max}} dE N\PE\exp[-E]}  \\&=\int_{E_{\rm min}}^{E_{\rm max}} dE \exp\left[S(E)-E\right].
\end{align}
where $S(E)=\ln N +\ln\PE$ is called the micro-canonical entropy in statistical mechanics. This integral can be estimated using a Laplace approximation:
\begin{align}\label{eq:saddle}
     T\propto \exp\left[S(E_m)-E_m\right],
\end{align}  
where $E_m$ maximizes the term inside the exponential. It satisfies the saddle-point condition $E_m=E^*$, where $E^*$ is defined by $S'(E^*)=1$, if that condition can be met within the range, i.e. if $S(E^*)>0$ or equivalently \add{if $S'(E_{\rm min})>1$}. Otherwise, the maximum value of $S(E)-E$ is reached at the lower bound of the range, $E_m=E_{\rm min}$. These two cases are illustrated geometrically in Fig.\,\ref{fig:rem}(a).

This implies two regimes for the quenched average that are picked based on the shape of $\PE$:
\begin{equation}\label{eq:qav}
    \mathcal{F}= \begin{cases}
        E^* -S(E^*) &\textrm{if }N>1/\mathcal{P}(E^*)\\
        E_{\rm min}
        &\textrm{otherwise.}
    \end{cases}
\end{equation}
Note that we have dropped terms of order $\ln\delta E$ consistent with the limit of the saddle-point approximation.
This transition is interpreted as a condensation transition between a high entropy state, where many antibodies of energy $E^*$ contribute to the titer $ T$, and a frozen state, dominated by a few antibodies with energy values around $E_{\rm min}$. These two regimes are also illustrated in Fig.\,\ref{fig:rem}(a).

To further highlight the difference between these regimes, we define $P_b(E)\propto \PE e^{-E}$ as the distribution of energy values conditioned on being bound to the virus, restricted to the range of observed antibodies $E\in(E_{\rm min},E_{\rm max})$.
We see in Fig.\,\ref{fig:rem}(b) how, in the non-condensed regime, the mode of this distribution is defined by a broad peak corresponding to many antibodies at $E^*$. In the condensed regime, binding is dominated by antibodies at $E_{\rm min}$.

\begin{figure}
    \centering
    \includegraphics[width=\linewidth]{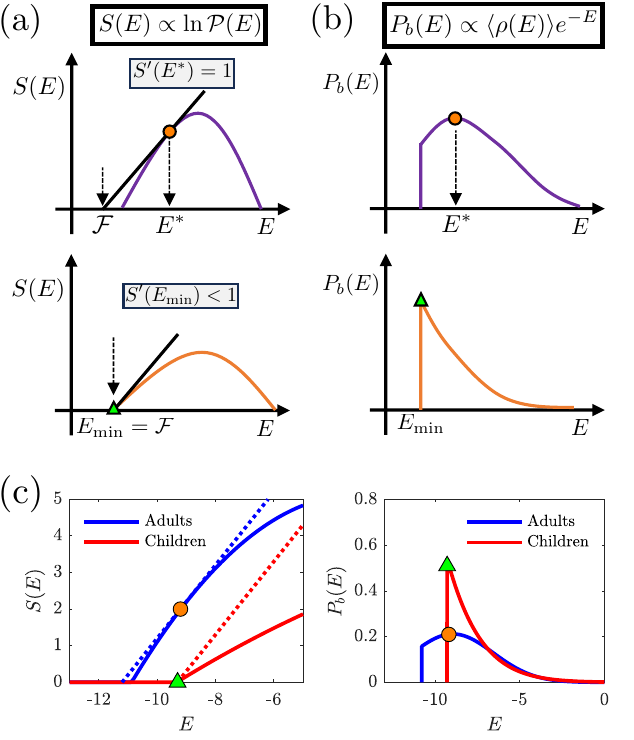}
    \caption{\textbf{Random Energy Model predicts two routes to sera neutralization ---} (a) Eq.\eqref{eq:qav} predicts two regimes for $\langle \mathcal{F}\rangle$. This depends on two energy values: $E^*$, as defined through $S'(E^*)=1$, and $E_{\min}$, set through $\mathcal{P}(E_{\min})=N^{-1}$. If $E^*>E_{\min}$, as for the purple curve, then $ \mathcal{F}$ reflects an average of the underlying $\PE$. If $E^*<E_{\min}$, $\mathcal{F}$ is set by $E_{\min}$ as shown for the orange curve. (b) This is also reflected in the mode of $P_b(E)$, the distribution of energy values conditioned on being bound to the virus: for the purple curve, this is set by $E^*$, whereas the orange case is dominated by the strongest clonotype at $E_{\min}$. (c) We plot $S(E)$ and $P_b(E)$ for the $\PE$ extracted for the adult and child datasets in Fig.\,\ref{fig:nums}(c), concluding on distinct routes to neutralization between the two.}
    \label{fig:rem}
\end{figure}

So far we have discussed the typical or average value of $\mathcal{F}$.
However, the two regimes also suggest different distributions of $\mathcal{F}$. In the non-condensed phase, since many antibodies contribute to the sum defining $ T$, we expect the law of large numbers to apply, so that $\mathcal{F}$ is tighly distributed around $E^*-\ln N-\ln\mathcal{P}(E^*)$.
In the condensed phase, $T$ is dominated by a few antibodies at the minimal value $E_{\rm min}$. In the extreme case where only one antibody would dominate, we would expect its statistics to be governed by extreme value theory, and more specifically by a Gumbel distribution when $\PE$ is unbounded and falls off sufficiently fast enough at low $E$ \cite{Haan2006, Hansen2020}. {In contrast to classical extreme value theory, in our case there may be more than one antibody contributing. However, we show numerically in Appendix \ref{SIsec:gumbel} that the Gumbel distribution provides a good fit at high $\sigma$ in our equilibrium binding model.}

The theory predicts a cross-over from Gaussian to a Gumbel distribution when $N<1/\mathcal{P}(E^*)$.
For a Gaussian $\PE$, the condensation condition is equivalent to $\sigma>\sigma_c\equiv \sqrt{2\ln N}\approx 3.7$ for $N=1,000$.
This transition line is plotted in black on the $(\mu,\sigma)$ phase diagram of Fig.~\ref{fig:nums}(b).

Our analysis of the REM demonstrates that asymmetric distributions for $\mathcal{F}=-\ln {T}$ are a reflection of extreme value statistics. Specifically, when sera rely on only a few strong clonotypes (i.e.\,those with the best affinity or the most abundant) to neutralize a virus, we expect $\mathcal{F}$ to fluctuate within a cohort according to the extreme value statistics of the underlying $\PE$. When neutralization is a collective effort of many clonotypes, we should expect the central limit theorem to hold and fluctuations within a cohort to be Gaussian. 

\subsection{Post-vaccination adult sera neutralization is a collective effect, whereas children rely on strongest clonotypes}

The previous analysis outlines the existence of two regimes, whether neutralization is dominated by a few or many antibody types. The two datasets in Fig.~\ref{fig:data}(c) illustrate these two regimes. Looking where the inferred values of $\mu$ and $\sigma$ fall in the phase diagram of Fig.~\ref{fig:nums}(b) for each cohort, we observe that vaccinated adults are in the uncondensed phase, while children are in the condensed phase.

In Fig.\,\ref{fig:rem}(c), we plot both the entropy $S(E)=\ln N +\ln \PE$, and the distribution of energies of antibodies participating in neutralization,
$P_b(E)\propto \PE e^{-E}$, for the post-vaccinated adults (blue) and the children (red).
For a Gaussian $\PE$, we have $E_{\rm min}=\mu-\sigma\sqrt{2\ln N}$.
While $E_m=E^*>E_{\rm min}$ for the adults, denoted by the orange circle, for the children we observe $E_m=E_{\min}$, denoted by the green triangle. Looking then to the mode of $P_b(E)$, we observe that the energy $E$ most expressed among binders for the children sera is the lowest, $E_{\rm min}$, whereas this is not the case for the vaccinated adults.

We conclude that the two datasets studied illustrate distinct routes to neutralization: for post-vaccinated adults, many clonotypes contribute, leading to Gaussian statistics for $\mathcal{F}$ as predicted by the central limit theorem, whereas the children's sera rely on the strongest clonotypes dominate the response, leading to non-Gaussian, extreme value statistics.

\subsection{Reliance on strongest clonotypes is the rule, rather than the exception}

\begin{figure}
    \centering
    \includegraphics[width=\linewidth]{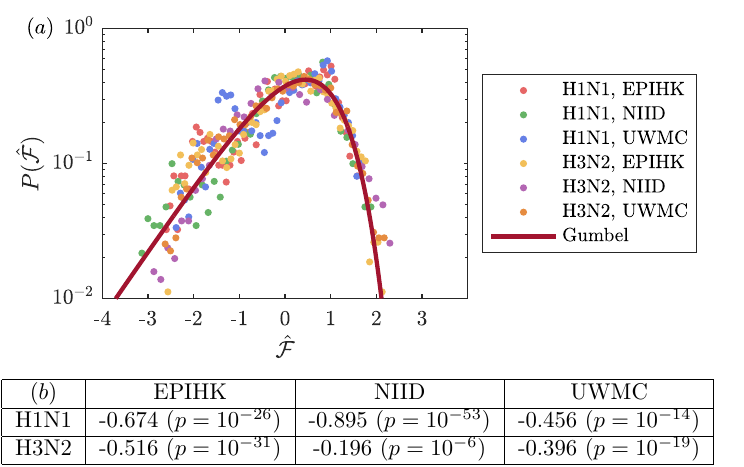}
    \caption{\textbf{Extreme value statistics are the rule rather than the exception ---} (a) The rescaled titers $\mathcal{F}=-\ln T$ from 6 different datasets of Ref.~\cite{Kikawa2025b} are plotted. For H1N1 and H3N2 viruses against sera from Hong Kong (EPIHK), Japan (NIID) and USA (UWMC), the titers collapse on to a Gumbel distribution, a signature of extreme value statistics. (b) For each dataset, the skewness is significantly different from zero: for D'Agostino's statistical test for zero skew, no dataset obtains a $p$-value greater than $10^{-6}$.  }
    \label{fig:rule}
\end{figure}

One may ask whether one of the two neutralization regimes is more common than the other. For example, is this driven by the cohort's age ranges, recent vaccination history or some other feature that distinguishes the two datasets? 
To address this question, we remark that the experimental approach detailed above has subsequently been used to measure neutralization titers for a broader range of sera against flu virus strains from 2024-25. This new dataset, first detailed in Ref.~\cite{Kikawa2025b}, further distinguishes between H1N1 and H3N2 viruses. It includes neutralization titers measured for predominantly adult sera taken from Hong Kong (EPI-HK Study at University of Hong Kong, $n=42$, ages 10-79), Japan (National Institute of Infectious Disease (NIID), $n=55$, ages 21-105) and the United States (University of Washington Medical Center (UWMC), $n=44$, ages 21-66) measured against H1N1 and H3N2 flu virus strains from 2024-25. (In analyzing the NIID data, we omit 7 post-vaccination follow up titers.) Separating the data by virus strain (H1N1/H3N2) and by origin of the sera, we identify 6 datasets. The vaccination status of the adults contributing these sera was largely unknown: only in the EPI-HK dataset was it known that 19/42 patients had received the 2024/25 Northern Hemisphere influenza vaccine and a twentieth patient had PCR-confirmed influenza virus infection identified in-house within 182 days of the serum collection.

For each of the six datasets, we repeat the same analysis as above to obtain distributions for $\mathcal{F}$ of the sera: we re-scale the titers as described in Eq.\,\eqref{eq:rescaleT} and again define the effective binding energies $\mathcal{F}=-\ln{T}$. We then compare the centered and normalized distributions of \add{$\hat{\mathcal{F}}=(\mathcal{F}-\langle \mathcal{F}\rangle)/\sqrt{\mathrm{Var}\mathcal{F}}$} across the datasets (Fig.~\ref{fig:rule}(a)). Each dataset fails D'Agostino's statistical test for zero skew, with no $p$ value greater than $10^{-6}$, indicating that fluctuations in log-titers across sera are non-Gaussian (Fig.~\ref{fig:rule}(b)).

Our previous analysis suggests that these non-Gaussian statistics may be caused by the dominance of a few strongest binders. If this is true, we should expect the distributions in Fig.~\ref{fig:rule}(a) to all agree with the Gumbel distribution from extreme value theory.
Fig.~\ref{fig:rule}(a) shows that this is indeed the case. This collapse provides quantitative evidence that neutralization within each cohort is controlled by a small number of the strongest clonotypes. It appears that this route to neutralizing a virus is the rule, rather than the exception, across the datasets studied.

\subsection{{Dynamics after vaccination: \add{selective activation of memory B cells} tightens $\PE$, {broadening the response}}}

\begin{figure}
    \centering
    \includegraphics[width=\linewidth]{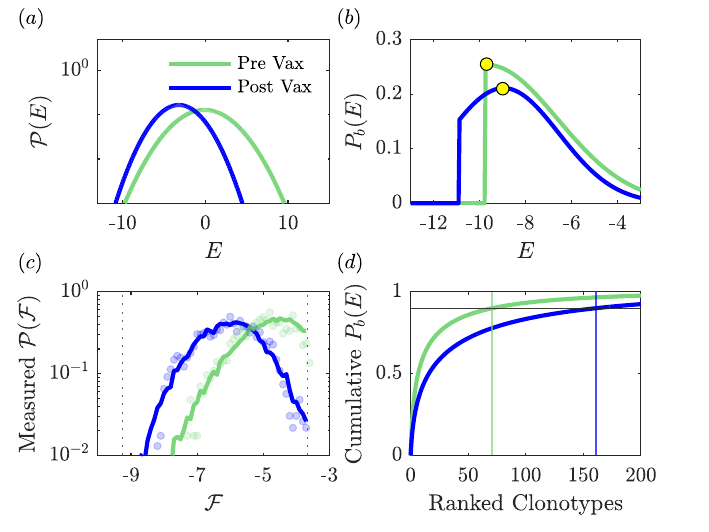}
    \caption{\textbf{Titer statistics and antibody affinities around vaccination ---} We compare the extracted $\PE$ from neutralization titers for adults pre- (day 0) and post- (day 28) vaccination. Post-vaccination data is replotted from Fig.~\ref{fig:nums}(c-d). (a) The distribution of affinities shifts to the left (stronger binders) and narrows due to the vaccine-elicited response. (b) This change in $\PE$ lowers the burden on the strongest binder, resulting in (c) Gaussian statistics post-vaccination. (d) Before vaccination, 90\% of viral binding is due to the strongest 71 clonotypes; after vaccination, this figure rises to 162 clonotypes, demonstrating an increasingly collective immune response.}
    \label{fig:prepost}
\end{figure}

So far, we have considered 8 datasets in this work, with 7 generating asymmetric titer distributions. The Gaussian statistics for $\mathcal{F}$ observed in Fig.\,\ref{fig:data}(c) for post-vaccination adults is an outlier. We reason that these sera were exceptional. Recent vaccination against influenza ensured that these sera contained many strong clonotypes, making neutralization a collective effort. Our results suggest that this is not the case in the typical, not recently vaccinated human sera, even against common viruses such as influenza. 

To explore this result, we consider the dynamics of the titer distribution shortly before and after influenza vaccination. Specifically, we return to the dataset of Ref.~\cite{Kikawa2025} which also includes neutralization titers for adults measured before (day 0) receiving influenza vaccination, as well as the post-vaccination (day 28) data that we have already discussed. In Fig.~\ref{fig:prepost}, we perform the same analysis as in Fig.~\ref{fig:nums}(c-d) to extract $\PE$ for pre- and post-vaccination datasets. The pre-vaccination $\PE$ (in green) is fit with $\mu=-0.1$ and $\sigma=3.1$. The post-vaccination fit (in blue) is the same in Figs.~\ref{fig:nums} and \ref{fig:prepost}: $\mu=-3.2$ and $\sigma=2.4$. In Fig.~\ref{fig:prepost}(a), the fit of a Gaussian model to the data shows the effect of the vaccination is to shift $\PE$ to the left and narrow the distribution. Initially, the pre-vaccine immune response was dominated by the strongest binders (green curve in Fig.~\ref{fig:prepost}(b)). {Perhaps counterintuitively,} this {tightening} of $\PE$ due to vaccination ensures neutralization becomes a collective effort amongst clonotypes (blue curve), {effectively broadening the response by increasing antibody diversity}. For the two extracted $\PE$ distributions, we find quantitative agreement between $\mathcal{P}(\mathcal{F})$ from the titer data and our model in Fig.~\ref{fig:prepost}(c). 

\add{This tightening can be motivated in a few ways. In a primary immune response, it would arise due to affinity maturation. In the Supplementary Information, we propose a mechanistic model for this that produces a tightening of the distribution: it is an extension of a similar model for the evolution of affinities during affinity maturation with an arbitrary (non-diffusive) mutation Kernel that was recently proposed by DeWitt \textit{et al.}~\cite{DeWitt2025}.}

\add{However, the response in adults observed in the current work is almost certainly a result of a secondary immune response \cite{MacLean2025}. In this case, the change in $\PE$ from before to after vaccination is a result of selective activation of memory B cells: these cells undergo rapid clonal expansion and differentiate into antibody-secreting plasma cells. Mathematically, we can define such a selection function $\mathcal{S}(E)$ such that the distributions can be related through \begin{equation}\mathcal{P}_{\rm post}(E)= \frac{1}{\mathcal{Z}}\mathcal{P}_{\rm pre}(E) e^{\mathcal{S}(E)}\end{equation}where $\mathcal{Z}$ enforces normalization. }

\add{To observe tightening of the distribution, we require that the selection function $\mathcal{S}(E)$ is concave: this non-linear selection is consistent with the idea that B-cell activation is not driven by equilibrium binding alone, rather it is a consequence of cooperative signaling and competition for T-cell help. Concavity is also indicative of diminishing returns: once antigen binding is sufficiently strong to reliably activate a memory B cell and secure T-cell help, further increases in affinity provide progressively smaller fitness advantages. A schematic choice which maps $\mathcal{P}_{\rm pre}(E)$ to $\mathcal{P}_{\rm post}(E)$ is $\mathcal{S}(E)=-aE-bE^2$, where $b>0$ imposes concavity. For the fitted distributions from Fig.~\ref{fig:prepost}, we solve for $a$ and $b$ that map from one distribution to the other, finding $a=0.545$ and $b=0.035$. }

\subsection{Effective number of antibodies contributing to neutralization}

Finally, we ask how many antibody clonotypes define the immune response in our model. 
We first estimate the fraction of viruses bound by the top $K$ antibodies:
the best $K$ binders to the virus have binding energies $E<E_K$ where $E_K$ is obtained by inverting $K = \int_{-\infty}^{E_K}\rho(E')dE'$; the fraction of bound virus due to these top $K$ clonotypes is then given by $\int_{-\infty}^{E_k} P_b(E')dE'\leq 1$. We then plot this fraction as a function of $K$, and ask how large $K$ should be to account for $90\%$ of the bound virus (Fig.~\ref{fig:prepost}(d)). In pre-vaccinated sera, this is achieved with 71 clonotypes (green vertical line), whereas this same coverage after vaccination is achieved with 162 clonotypes (blue vertical line). We remark that while the cutoff of $90\%$ is somewhat arbitrary, the $2.3$-fold increase in number of relevant clonotypes is robust: for a cutoff of $50\%$, the fold change is 2.4; for $95\%$, it is 2.3.

We compare these estimates to data from Lee et al.~\cite{Lee2016}, where a combination of B-cell repertoire sequencing, proteomic identification of serum antibodies, and expression of antibody clonotypes experimentally quantified the structure of the polyclonal response to 3 strains of influenza constituting the trivalent seasonal influenza vaccine, in 4 patients. Against the H1N1 component, responses were initially defined by 29-44 clonotypes, whereas after vaccination, the study identified 55-118 relevant clonotypes for each donor. Against the H3N2 component, results were more varied: one donor's pre-vaccination immune response consisted of only 6 clonotypes, while other's ranged from 14-99. After vaccination, 40-116 clonotypes were identified. These numbers, and in particular the fold change from pre- to post-vaccination, are consistent with our theoretical prediction. Note that our model does not track the specific identities of each clonotype: in Lee et al.~\cite{Lee2016}, clonotypes appearing in the pre-vaccinated sera accounted for only 60\% of those in post-vaccination sera.

\section{Discussion}

We have illustrated how fluctuations in neutralization titers within a cohort reveal the structure of antibody immune responses. This structure can inform us  about immune robustness \cite{Mayer2015,Lee2019b, Schnaack2021, Greaney2021b, MunozAlia2021, Chardes2022} and {help us predict evolutionary trajectories during virus-immune coevolution \cite{Mayer2016b, Lassig2017, Lassig2020, Marchi2021, Chardes2023}}: narrow responses may be susceptible to viral escape through single-point mutations \cite{Starr2020, Greaney2021a, Cao2022, Cao2022b}, limiting the durability and breadth of protection. In contrast, broadly polyclonal responses provide a more resilient immunity, as viral escape would potentially require coordinated multiple mutations across epitopes \cite{Scheid2009, Yu2022}. {We observe that in the majority of cases, these titer fluctuations are well described by extreme value statistics. This implies that the immune response is generally dominated by a few of the strongest antibody clonotypes, consistent with molecular-level studies \cite{Wine2013, Lee2016}, {suggesting a narrow neutralization capability that is not robust to escape variants}.}

Only for the neutralization titers of post-vaccinated adults were the fluctuations not captured by extreme value statistics. We argued that this was due to the antibody repertoire being strengthened through \add{selective B cell activation} 
meaning the neutralization of the virus became a collective effort across many high-affinity antibody clonotypes. {This is in contrast to what might be observed from clonal bursting in a single germinal center, where a single high-affinity clonotype is expanded to provide a strong immune response \cite{Victora2022,DeWitt2025}. Our results are consistent with the idea of ``bet-hedging" during affinity maturation \cite{Zhang2014, DiNiro2015, Kuraoka2016, Chardes2022, Hagglof2023, Chen2023, Sprumont2023, Yang2023, Schiepers2024}, where immune responses tend to expand multiple high-affinity clonotypes to maintain diversity in epitope specificity {through a multiplicity of germinal center reactions}.}

\add{In the Supplementary Information, we} proposed a mechanistic model for how affinity maturation could also achieve this. Specifically, we showed how somatic hypermutation, affinity-based selection and the replacement of low-affinity antibody lineages by higher-affinity ones can lead to a tightening of $\PE$, the distribution of antibody affinities and serum composition. Without this turnover, the distribution $\PE$ would broaden due to both somatic hypermutation \cite{Pae2025} and selection pressures. Lineages that have high affinity, or improve it through random mutations, expand, while others are suppressed. Lineage turnover is crucial to realize the observed tightening of $\PE$: it is a constraint that future models of affinity maturation should consider: \add{We remark that this turnover is necessary as an addition to the affinity maturation model of Ref.\,\cite{DeWitt2025} in order to recapitulate the observed narrowing of the distribution.}

{The observed tightening of $\PE$ could also be tested experimentally: antibody-level analysis as in Refs.\,\cite{Wine2013, Lee2016} could be used to illustrate how the real distribution of affinities evolves throughout the immune response. As discussed in the Introduction, there are limitations to these approaches, but we expect that a tightening of $\PE$ should be observable even from an incomplete sampling of the immune repertoire {specific for the antigens of interest}. Also, more high-throughput titer data would allow us to extend our analysis. For example, titers measured beyond 28 days post-vaccination would inform a model for how affinities evolve after the initial immune response tightens $\PE$, \add{resulting in a more diverse response}. This would allow us to estimate the timescale over which this diversity is lost, and the regime of extreme values dominated by a few antibodies is recovered. \add{Titer data after repeat vaccinations to seasonal strains \cite{Cowling2024} or in response to consecutive infection \cite{VanBeek2022, Sun2026} 
  could be used to understand how the complex dynamics of immune memory may affect our conclusions.} {For instance we expect imprinting, or original antigenic sin, to maintain dominant antibodies in their dominant position even after vaccination \cite{Cobey2017, Lewnard2018, Dugan2020}, which could explain why titers measured long after vaccination are dominated by a few antibodies and thus well described by extreme value statistics.}}

We remark that one drawback of looking at higher-order moments (e.g.\,the skewness) of experimentally-measured distributions is that these moments can fluctuate wildly in small datasets. Careful treatment is required to measure these moments such that they truly reflect a property of the whole underlying distribution. We have demonstrated here that the high-throughput approaches developed and employed in Refs.~\cite{Kikawa2025} and \cite{Kikawa2025b} for measuring neutralization titers to a broad range of viral strains are sufficient to achieve this. We expect that future titer data measured from these techniques will produce data that is ripe for similar analysis.

{We found that the empirical statistics of log-titers are often consistent with a Gumbel distribution}, which {suggests} that the underlying $\mathcal{P}(E)$ has tails that decay {faster than an exponential}, such as for a Gaussian distribution.
\add{This in turn implies that the real distribution of binding affinities across sera is unlikely to be fat-tailed or saturating near a maximum-affinity value \cite{Batista1998}. Recent work proposed an exponential distribution for the germline affinity naive B cells \cite{Yang2023}: the observed faster-than-exponential decay for $\PE$ (i.e. for mature B cells) suggests that maturation may function to reduce variability in affinities, which is consistent with the discussion above.}

{The Gumbel distribution describes the extreme value statistics for a Gaussian distribution, but its connection to the partition function, the sum in Eq.\,\eqref{eq:fzm1}, remains to be clarified. For a Gaussian $\PE$, Eq.\,\eqref{eq:fzm1} is a sum of log-normal distributed random variables. In the limit of large $N$, the central limit theorem implies that the statistics of the partition function should be Gaussian across realizations. As we discuss in Eq.\,\eqref{eq:qav}, a departure from Gaussian statistics appears at {finite} $N$, in particular for $N<\exp(\sigma^2/2)$ for a Gaussian $\PE$ with variance $\sigma^2$. While we have shown numerically that Gumbel statistics arise when we are deep in this regime (see Appendix \ref{SIsec:gumbel}), a mathematical characterization of this crossover remains an open problem.} 

{More generally, our analysis was based on a generic equilibrium binding model for antibody-virus interactions. We have demonstrated that this model can quantitatively capture titer statistics from data for influenza. We believe the generality of our binding model enables us to extend it directly to study different pathogens. Indeed, recent work has applied the same density of states formalism in a pathogen-agnostic approach to describe the initial expansion of B cell repertoires upon infection \cite{MoranTovar2024}. We expect our approach to be applicable to other fast-evolving respiratory viruses like SARS-CoV-2.}

{Another interesting open question is to characterize immune responses to less common pathogens in a similar manner. We have showed here that even for common pathogens like influenza, the baseline immune response is dominated by extreme binders. We expect this domination to be even more significant in responses to rare pathogens, further highlighting the relevance of extreme value statistics for describing neutralization titers. On the modeling side, including imperfect or cooperative neutralization \cite{Mouquet2010, Einav2020}, correlations between antibody affinities \cite{Derrida1985} or antibody-antibody interactions at binding epitopes each provide promising avenues towards more realistic models of the immune response at the level of antibody clonotypes. }

\section*{Acknowledgements}
We thank the referees for their helpful comments and suggestions. This study was supported by  the European Research Council Proof of Concept grant no 101185627, and the Agence Nationale de la Recherche grant no ANR-19-CE45-0018 “RESP-REP” and ANR-24-CE45-7957 "WILDTYPES",  Fondation Bettencourt Schueller, Foundation pour la Recherche Medicale grant Team Project EQU202503019997 and the CZI Theory Initiative grant.

\bibliographystyle{pnas}

\onecolumngrid
\renewcommand{\thefigure}{S\arabic{figure}}
\renewcommand{\thetable}{S\arabic{table}}
\setcounter{figure}{0}
\setcounter{table}{0}

\appendix

\section{\ha{Data Availability}}

The data and MATLAB scripts used to plot all figures in this work are available at:
\url{https://github.com/statbiophys/NeutralizationREM_Figures/tree/main}.

\section{Derivation and numerical analysis of Eq.~\eqref{eq:pfreeM}}\label{sec:SIderiv}

We explain here the form of Eq.~\eqref{eq:pfreeM} which determines the probability that \add{a tile on a viral particle} is unbound at antibody concentration $c$. We denote this probability by $p_{\rm free}(c)$. Our derivation is valid in the limit where the concentration of the virus is limiting, meaning the concentration of antibodies is approximately equal to the concentration of free (unbound) antibodies. As a consequence, we can neglect the effect of the concentration of the virus, and treat each \add{tile} as if it was on its own. We show below that  $p_{\rm free}(c)$ can be defined through the dissociation constants of the constituent antibodies of the sera to the virus. Before arriving at the full result Eq.~\eqref{eq:pfreeM}, we build up intuition by deriving $p_{\rm free}(c)$ for simple cases. 

\subsection{Single epitope, $M=1$}

First consider the case of a monoclonal antibody of concentration $c$ which binds to a single epitope \add{on a tile} of a viral particle with a dissociation constant $K_D$. In a volume $V$, the total concentration of viral particle \add{tiles} is $1/V$. The concentration of free viral particle \add{tiles} is [free virus] $=p_{\rm free}/V$, and that of bound \add{tiles} is [virus-antibody complex]$\add{=(1-p_{\rm free})/V}$. The definition of the equilibrium constant,
\begin{equation}
  K_D=\frac{\textrm{[free virus][antibody}\textrm{]}}{\textrm{[virus-antibody complex}\textrm{]}}
\end{equation}
imposes the ratio of these two quantities must be equal to
\begin{equation}
 \frac{1-p_{\rm free}(c)}{p_{\rm free}(c)} =\frac{c}{K_D}.
\end{equation}
This leads to:
\begin{equation}\label{sieq:first}
  p_{\rm free} (c)
  = \frac{1}{1 + \frac{c}{K_D}}.
\end{equation}
\add{thus the concentration at which a given percentage of tiles is bound scales linearly with $K_D$. }

Now consider $N$ antibody clonotypes which compete for the same epitope. Antibody $a$ is at concentration $f_ac$. Call $p_a$ the probability that an epitope is bound by antibody $a$, and $p_{\rm free}=1-\sum_{a=1}^Np_a$. The equilibrium condition for each antibody,
\begin{equation}
  K_D^a=\frac{\textrm{[free virus][antibody }a\textrm{]}}{\textrm{[virus bound by }a\textrm{]}},
\end{equation}
dictates
\begin{equation}
  \frac{p_a}{p_{\rm free}}=\frac{f_ac}{K_D^a}.
\end{equation}
Summing over $a$ gives:
\begin{equation}
  \frac{1-p_{\rm free}}{p_{\rm free}}=c\sum_{a=1}^N\frac{f_a}{K_D^a},
  \end{equation}
  from which we derive the fraction of free viral \add{tiles} as:
\begin{equation}\label{sieq:m1}
p_{\rm free}(c) = \frac{1}{1 + \sum_{a} \frac{f_ac}{K^a_D}} = \frac{1}{1 +  \frac{c}{\mathcal{K}_D}},
\end{equation}
where $\mathcal{K}_D^{-1}=\sum_a f_a {K^a_D}^{-1} $ is the \textit{harmonic mean} of the constituent $\{K_D^i\}$. The polyclonal sera behaves as a monoclonal antibody with dissociation constant $\mathcal{K}_D$. 

\subsection{Multiple epitopes, $M>1$}\label{sec:SImtopes}
To arrive at Eq.~\eqref{eq:pfreeM}, we want to consider multiple binding epitopes on \add{each tile of} the viral particle. We denote by $M$ the number of epitopes. We make two simplifying assumptions: (i) each antibody clonotype only binds to one epitope on the virus and (ii) binding at distinct epitopes is independent. We define $e_a$ as the epitope which clonotype $a$ binds to and $F_e=\sum_{a:e_a=e}f_a\leq 1$ the fraction of antibody clonotypes that bind to epitope $e$. 

Each viral \add{tile} is free if each of its epitope is unbound. Independent binding then implies that
$p_{\rm free}(c)$ is just a product of terms like the one appearing in Eq.~\eqref{sieq:m1}:
\begin{align}\label{sieq:pfreeM}
p_{\rm free} (c)&= \prod_{e=1}^M\left(\frac{1}{1 + \sum\limits_{a:e_a=e}\frac{c f_a}{K^a_D}}\right)= \prod_{e=1}^M\left(\frac{1}{1 + \frac{c F_e}{\mathcal{K}_D^{e}}}\right)
\end{align}
with $\mathcal{K}_D^{e}$ the harmonic mean of all dissociation constants for epitope $e$. 

\add{For $M>1$, Eq.\,\eqref{sieq:pfreeM} now appears as a complicated product of terms, unlike the case for $M=1$. However, we argue that the harmonic mean of all dissociation constants, namely $\mathcal{K}_D^{-1}=\sum_e F_e /{\mathcal{K}_D^{e}}=\sum_a f_a /{K^a_D}$, can still be used to quantify binding in this model. To see this, it is instructive to consider two extreme cases. First, in the case where there are many epitopes but one epitope (say $e=1$) is heavily targeted by antibody clonotypes (either through a large fraction $F_1$ or through stronger binding through $\mathcal{K}_D^1$), we expect the single epitope picture to hold. Specifically, assuming that $F_1/K_D^1\gg F_e/K_D^e$ for all $e>1$, we see that $\mathcal{K}_D\approx \mathcal{K}_D^1/F_1$ and, upon expanding the product in Eq.\,\eqref{sieq:pfreeM}, that  \begin{equation}p_{\rm free}(c)=\frac{1}{1+\frac{cF_1}{\mathcal{K}_D^1}+\dots}\approx \frac{1}{1+\frac{c}{\mathcal{K}_D}}\end{equation}where the ellipsis contains terms of order $F_e/K_D^e$ which are small compared to the first two terms in the denominator.  }

\add{A second extreme example is the limit where many epitopes are equally targeted: $F_e= 1/M$ and $K_D$ is constant across clonotypes (which in turn implies $\mathcal{K}_D=K_D$). In this case, the expansion approximates the exponential function \begin{equation}p_{\rm free}(c)=\left(\frac{1}{1+\frac{c}{M\mathcal{K}_D}}\right)^M\approx e^{-c/\mathcal{K}_D}\end{equation}and hence we again see that the concentration at which a given percentage of tiles is bound must scale linearly with $\mathcal{K}_D$.}

\begin{figure*}[t]
    \includegraphics[width=\linewidth]{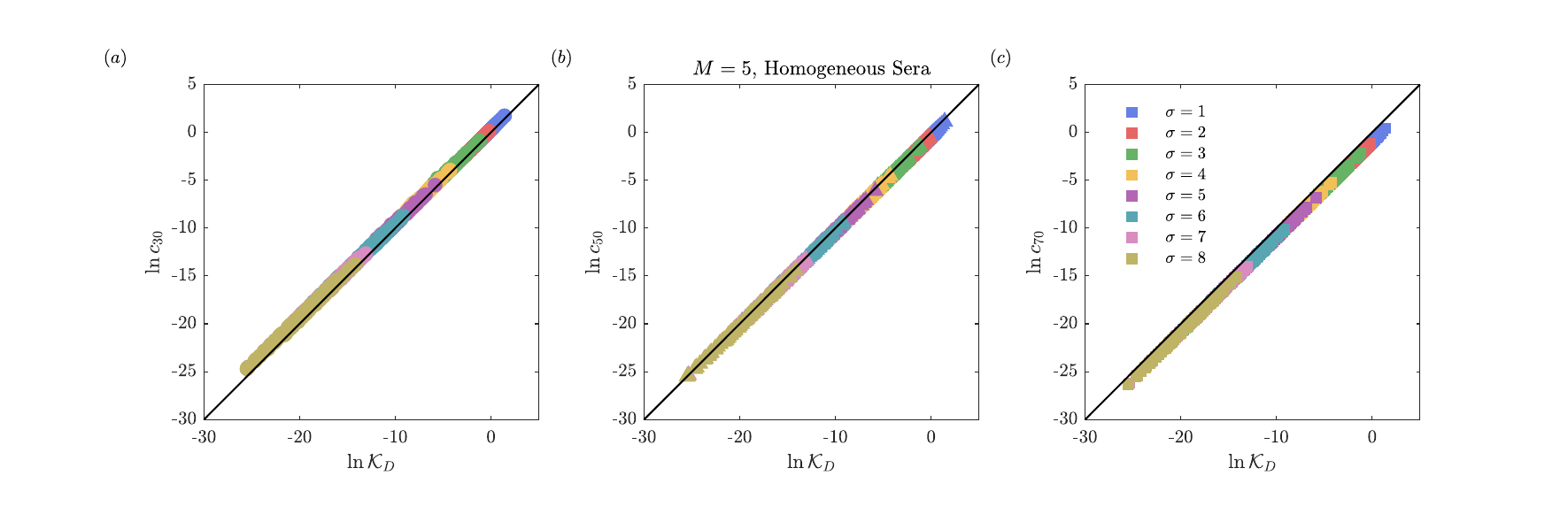}
    \caption{\add{\textbf{Analysis of Eq.~\eqref{sieq:pfreeM} for $M=5$ and $f_a\equiv 1/N$ ---} We observe a linear relationship between $\log \mathcal{K}_D$ and (a) $\ln c_{30}$, (b) $\ln c_{50}$ and (c) $\ln c_{70}$ across a wide range of $\sigma$ (indicated by colours). The black line denotes $x=y$. This fit justifies the scaling relation $-\ln T\propto \ln \mathcal{K}_D$ (used in the main text to derive Eq.\,\eqref{eq:fzm1}) for case of homogeneous sera, $f_a\equiv1/N$.}}  
    \label{fig:mepitopes5}
\end{figure*}

\begin{figure*}[t]
    \includegraphics[width=\linewidth]{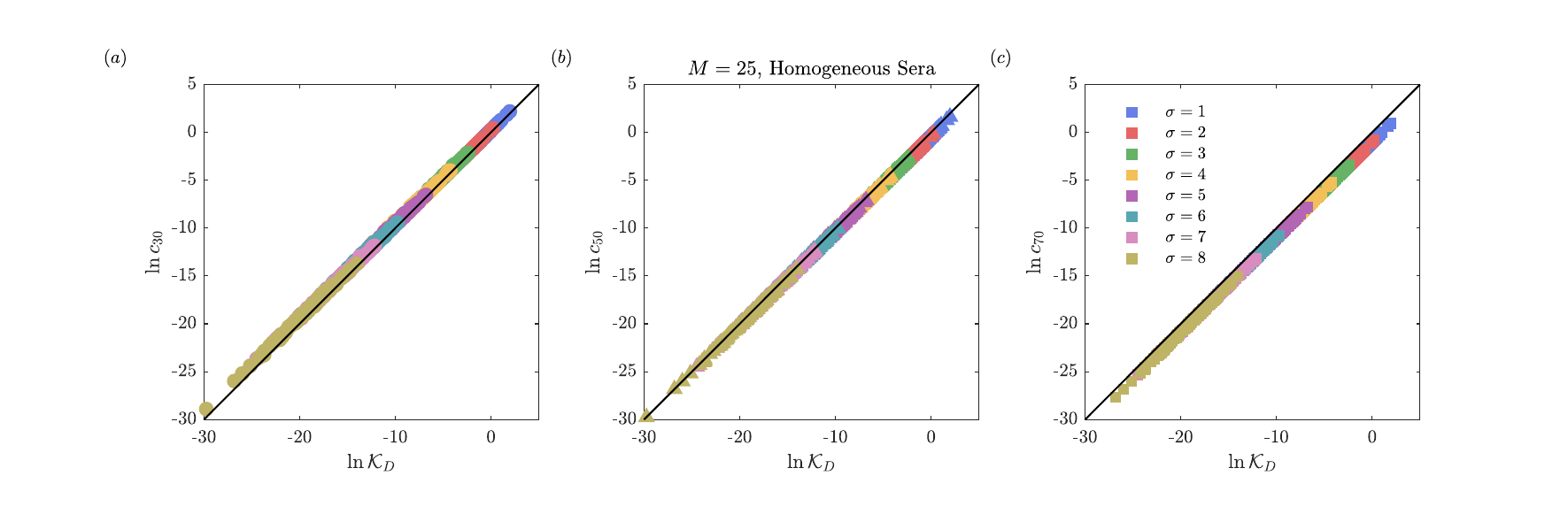}
    \caption{\add{\textbf{Analysis of Eq.~\eqref{sieq:pfreeM} for $M=25$ and $f_a\equiv 1/N$ ---} We confirm the same linear scaling of Fig.\,\ref{fig:mepitopes5} holds for a larger number of epitopes, $M=25$.}}
    \label{fig:mepitopes25}
\end{figure*}

\begin{figure*}[t]
    \includegraphics[width=\linewidth]{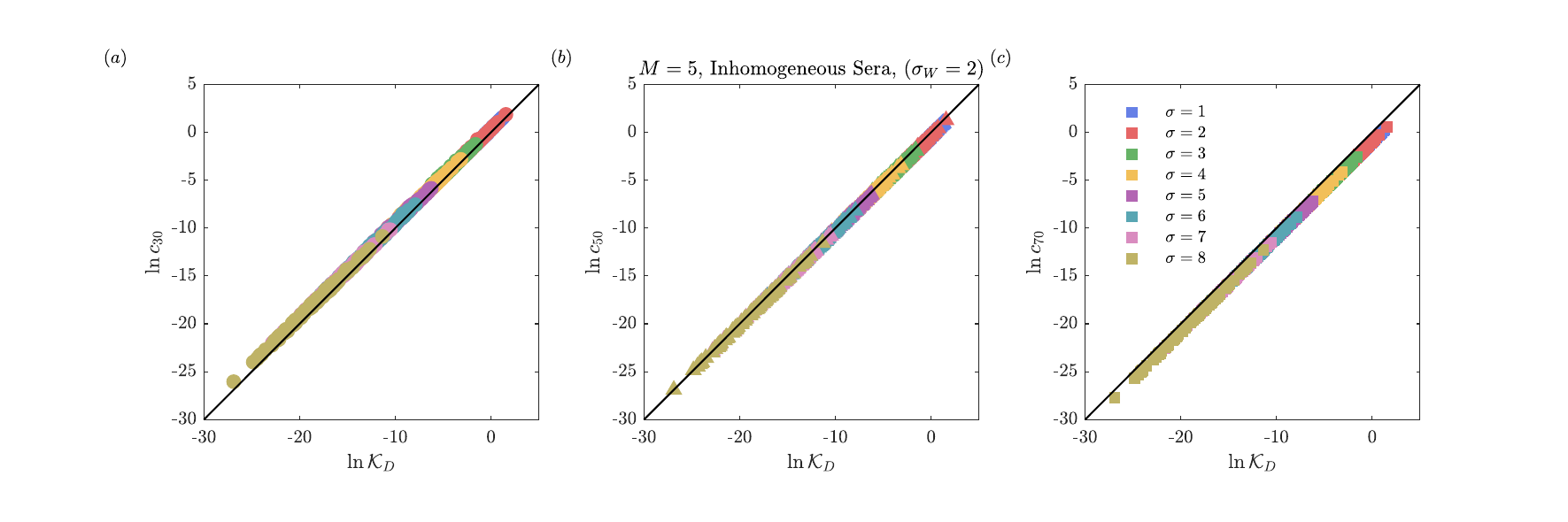}
    \caption{\add{\textbf{Analysis of Eq.~\eqref{sieq:pfreeM} for $M=5$ and unequal $f_a$ ---} We confirm the same linear scaling of Fig.\,\ref{fig:mepitopes5} holds when the serum fractions are unequal (here with $\sigma_W=2$). }}
    \label{fig:mepitopes5I}
\end{figure*}

We now show that it {\TM approximately} holds in general, starting with homogeneous sera. Specifically, we investigate how $c_{30}$ (concentration at 30\% bound tiles), $c_{50}$ and $c_{70}$ scale with $\mathcal{K}_D$ in the full result Eq.~\eqref{sieq:pfreeM} through numerical analysis.

We consider a log-normal distribution for the dissociation constants $\{K_D^a\}$ (or equivalently Gaussian $\mathcal{P}(\phi)$ for $\phi_a\propto\log K_D^a$). We sample $N=1,000$ values for dissociation constants, repeating this $N_s=10^4$ times. As in the main text, we model variability between sera by saying that each serum is described by a Gaussian distribution $\mathcal{P}(\phi_a)$ of antibody affinites with a serum-independent variance $\sigma^2$ but a serum-dependent mean $\mu_s$, which is itself normally distributed with mean $\mu$ and variance $\sigma_1^2$ across individuals. As in the main text, we set $\sigma_1=0.9$ and further set $\mu=0$ for simplicity. We assign each $K_D^a$ an epitope $e_a$ sampled from $\{1, \dots, M\}$ with equal probability, $1/M$. We first assume a homogeneous serum, $f_a\equiv 1/N$.

For each realization we calculate the three concentrations (from Eq.~\eqref{sieq:pfreeM}) and $\mathcal{K}_D$ (the harmonic mean). The natural log of these quantities is then plotted against each other for different values of $\sigma\in\{1, 2, \dots,8\}$ (indicated by colors) for $M=5$ epitopes in Figure \ref{fig:mepitopes5} and $M=25$ epitopes in Figure \ref{fig:mepitopes25}

Finally, we show the effect of different serum fractions $f_a\neq 1/N$ between clonotypes. We perform the same numerical analysis as above, but now with $M=5$ fixed. We assign each of the $N$ clonotypes a weight $w_a$ from a log-normal distribution such that $\ln(w_a)$ is distributed like a Gaussian with zero mean and standard deviation $\sigma_W$. To obtain the serum fractions, we then normalize the weights such that they sum to 1: $f_a = w_a / \sum_{a'}w_{a'}.$ \add{We make the choice here that each epitope is targeted equally. We expect that the case where a handful of epitopes $M’ < M$ are heavily targeted is qualitatively similar to the case where $M’$ epitopes are targeted equally, hence why we consider only homogeneous epitope targeting. }In Fig.\,\ref{fig:mepitopes5I}, we demonstrate the results Eqs.~\eqref{eq:FT} and \eqref{eq:fzm1} accurately describe the case of inhomogeneous sera for $\sigma_W =2$.

\subsection{{\TM Validity of $\mathcal{F}=\ln \mathcal{K}_d+\textrm{const}$ for cooperative antibodies}}

\add{We now return to the main result for our model in the main text, namely Eq.\,\eqref{eq:FT} connecting the titer $T$ to the composition of the serum {\TM when antibody may bind cooperatively}.}
{\TM We model the binding state of each viral tile $i=1,\ldots,Q$ with a ``spin'' taking value $\sigma_i=+1$ if the tile is bound, and -1 if it is unbound. In general, the neutralization probability is a function of the collective binding state of all tiles: $P_{\rm neut}(\sigma_1,\ldots)$.

In absence of cooperativity between antibodies, the Boltzmann weights of the bound and unbound states are given by \ha{$w_+\propto 1-p_{\rm free}(c)$ and $w_-\propto p_{\rm free}(c)$}, resulting in the effective free Hamiltonian:
  \begin{equation}
    H_{\rm free}(\sigma_1,\ldots)=-h\sum_{i=1}^Q \sigma_i
  \end{equation}
  with
  \begin{equation}
h=\ha{\frac{k_BT}{2}\ln\frac{w_+}{w_-}.}
  \end{equation}
$h$ depends on $c$ only through $c/\mathcal{K}_d$, as we argued in the previous section. In the case of a single epitope, it reads:
  \begin{equation}
\ha{h=\frac{k_BT}{2}\ln\frac{c}{\mathcal{K}_d}.}
  \end{equation}

  In addition, bound antibodies may interact in a way that is independent of their type, through a generic interaction Hamiltonian $H_{\rm int}(\sigma_1,\ldots, \sigma_Q)$. The equilibrium distribution of binding states is then given by Bolztmann's distribution:
  \begin{equation}
    P_{\rm eq}(\sigma_1,\ldots)=\frac{1}{Z}\exp\left[-\frac{H_{\rm int}(\sigma_1,\ldots)+H_{\rm free}(\sigma_1,\ldots)}{k_BT}\right].
  \end{equation}
  The overall probability of neutralization, $\bar P_{\rm neut}$ is the average probability of neutralization $P_{\rm neut}(\sigma_1,\ldots)$ over all binding configurations:
  \begin{equation}
    \bar P_{\rm neut}=\sum_{\sigma_1,\ldots}P_{\rm eq}(\sigma_1,\ldots)P_{\rm neut}(\sigma_1,\ldots)=f\left(\frac{c}{\mathcal{K}_d}\right),
    \end{equation}
  which only depends on $c$ through $c/\mathcal{K}_d$.

The $c_{50}=c_0/T$ at which half-neutralization occurs is then given by $c_{50}=\mathcal{K}_df^{-1}(1/2)$.}

\section{Setting $\sigma_1$ from data} \label{SIsec:sigma1}

When comparing to data, we look to set $\sigma_1$ which captures the inherent variability in affinities between two sera in the same cohort. In the main text, we fit this using the variability in the post-vaccination adult sera. The argument was that when the variance across antibodies of $E$, $\sigma$, is below the condensation transition, the variability in $\mathcal{F}$ is largely dictated by $\sigma_1$. Since the post-vaccinated adults are the only dataset in this small-$\sigma$ regime, we use this to set $\sigma_1$. 

In Fig.\,\ref{fig:SIsigma1} we justify our claim that $\sigma_1$ sets the variability in $\mathcal{F}$ when $\sigma$ is small. We consider numerical realizations for three values of $\sigma_1$. At small $\sigma$, this standard deviation is approximately equal to $\sigma_1$ (denoted with dashed lines for the three cases) for all $\sigma_1$ values considered. At large $\sigma$, a second contribution linear in $\sigma$ emerges, as the system becomes condensed and is dominated by extreme events. For the post-vaccinated adult dataset, the standard deviation in $\mathcal{F}$ is 0.9 (denoted with the black horizontal line in Fig.\,\ref{fig:SIsigma1}), thus we choose this value for $\sigma_1$ to capture the implicit variability between sera throughout the manuscript. The inferred value of $\sigma=2.4$ for vaccinated adults places us in the regime where the contribution of $\sigma$ to $\mathrm{Var}(f)$ matters little, consistent with our assumption.

\begin{figure*}[t]
    \includegraphics{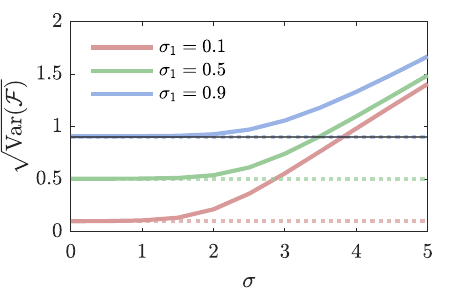}
    \caption{\textbf{Setting $\sigma_1$ from data ---} We use the post-vaccination adult dataset to set $\sigma_1$ in our work in the following way: we first remark that at small $\sigma$ the variability in $\mathcal{F}$ is set entirely by $\sigma_1$ (dashed lines, demonstrated here for three values of $\sigma_1$). Specifically, $\sqrt{\textrm{Var}{(\mathcal{F})}}\approx\sigma_1$. Since the post-vaccinated adult dataset is the only one which we fit with a small value for $\sigma$, we use the standard deviation for $\mathcal{F}$ measured for this data to set $\sigma_1=0.9$.}  
    \label{fig:SIsigma1}
\end{figure*}

\section{Gumbel statistics at large $\sigma$}\label{SIsec:gumbel}

Finally, we also argue for the presence of Gumbel extreme values statistics in the regime where neutralization is dominated by the strongest binders. This arises when the partition function, i.e. the sum in Eq.\,\eqref{eq:fzm1}, is dominated by the term from the smallest value for $E$. This would imply \add{$\mathcal{F} = -\ln T \sim -\ln \exp(-E_{\min}) = E_{\min}$}, thus $\mathcal{F}$ would be distributed according to the extreme value statistics of $\PE$. For our Gaussian $\PE$, we would thus expect Gumbel statistics. In practice, this is only exactly true when no other energy values contribute to the sum.

Here, we justify our claim of Gumbel statistics by referring to numerical realizations of our model at high variability $\sigma$, where we argue these extreme value statistics emerge. This is demonstrated in Fig.\,\ref{fig:SIgumbel} where at large $\sigma$ ($\sigma=8$) the distribution for $\hat{\mathcal{F}}\equiv (\mathcal{F}-\<\mathcal{F}\>)/\<(\mathcal{F}-\<\mathcal{F}\>)^2\>^{1/2}$, in blue, is accurately fit by a Gumbel distribution (dashed blue line).
{A mathematical characterization of the transition between these two regimes remains an open problem.}

\begin{figure*}[t]
    \includegraphics{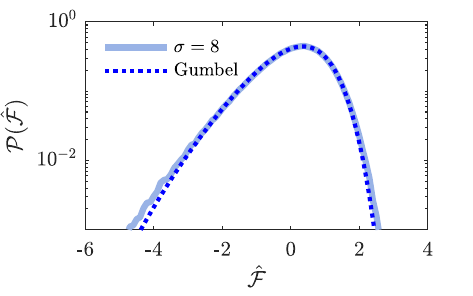}
    \caption{\textbf{Gumbel statistics at large $\sigma$ ---} We justify our claim that the extreme value statistics observed at large $\sigma$ in our model are of Gumbel form. In light blue are $\hat{F}$ statistics measured from realizations over $10^6$ sera ($\mu=0$ and $\sigma_1=0.9$). These are fit well by a Gumbel distribution (dashed blue line).
    }  

    \label{fig:SIgumbel}
\end{figure*}

\section{Mechanistic model for affinity maturation and tightening of $\PE$}

\begin{figure}
    \centering
    \includegraphics[width=0.5\linewidth]{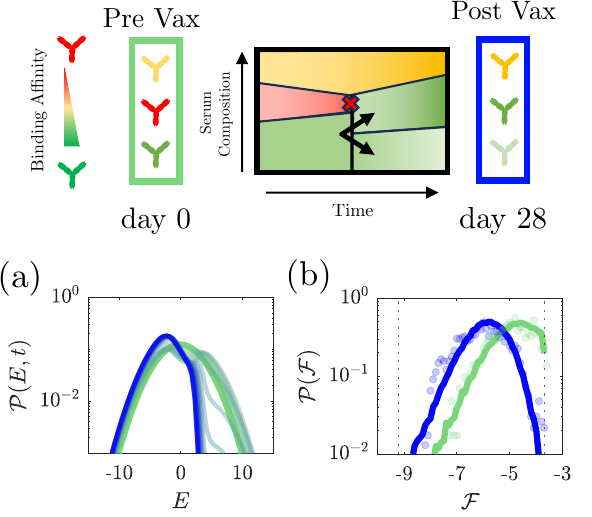}
    \caption{\textbf{Mechanistic model of affinity maturation captures $\PE$ dynamics across vaccination ---} \add{Schematic of the clonotype termination (red cross) and branching (black arrows) mechanism implemented in our mechanistic model. Gradients in color denote changes in affinity due to somatic hypermutation.} (a) We simulate the dynamics for the affinities (Eq\,\eqref{eq:dphi}) and serum fraction (Eq.\,\eqref{eq:df}) and plot the resulting distribution $\mathcal{P}(E,t)$, averaged over simulations. The dynamics result in a narrowing of the distribution over time due to the termination of low-affinity (and expansion of high-affinity) clonotypes. (b) The initial and final distributions for $\PE$ are then used to numerically sample a distribution for $\mathcal{F}$ (plotted with solid lines) which quantitatively agrees with the data (plotted with circular points). Note the data is the same as in Fig.\,\ref{fig:prepost}(c) but now using the simulation result in (a) as the input for $\PE$.  }
    \label{fig:am}
\end{figure}

In this section, we propose a mechanistic model to illustrate how this change in $\PE$ can occur through affinity maturation. We model the dynamics at the level of distinct B-cell lineages, and attribute to them the serum fraction and affinity of the antibodies they secrete. Following antigen exposure, the initial repertoire of clonotypes enters germinal centers (GCs) \cite{DeSilva2015, Tas2016}. During affinity maturation, each clonotype undergoes somatic hypermutation in the dark zone (DZ), resulting in a larger range of affinities. Upon transitioning to the light zone (LZ), clonotypes compete for antigen binding, where their probability of survival (and subsequent re-entry into the DZ) depends on their affinity $\phi_a$ and fraction $f_a$. The result of this iterative cycle of mutation and selection is the proliferation of high-affinity clonotypes, ultimately producing antibodies with enhanced specificity to the target antigen. 

We model affinity maturation here as a continuous time process in which the affinity and serum fraction of each clonotype evolve in the following way: the affinity $\phi_a(t)$ evolves as a symmetric random walk modeled with a Gaussian white noise as
\begin{equation}\label{eq:dphi}
\dot{\phi}_a(t) = \sqrt{2D}\eta_a(t)
\end{equation}
where $D$ denotes an effective diffusion coefficient for the walk, and $\eta_a(t)$ a unit uncorrelated Gaussian white noise. Larger diffusion corresponds to more rapid changes in affinity due to mutations.

The change in serum composition is driven by selective pressures acting on $f_a(t)$. We assume that the proliferation of clonotype $a$ is solely a function of its affinity $\phi_a(t)$, which leads to the following replicator dynamics for $f_a$:
\begin{equation}\label{eq:df}
\dot{f}_a(t) = f_a(g(\phi_a)-\bar{g}(t)),\quad \bar{g}(t) = \sum_a f_a(t) g(\phi_a(t)),
\end{equation}
where $g(\phi)$ captures how affinity drives selection in the LZ and $\bar{g}(t)$ ensures that $\sum_{a}f_a=1$ for all $a$ and $t$. Selection here favors high-affinity clonotypes, thus we expect $g(\phi)$ to be a non-increasing function of $\phi$. Below we parameterize this function as $g(\phi)=-r\tanh((\phi-\phi_c)/\alpha)$, where $\phi_c$ sets the affinity and $\alpha$ sets the sharpness of the cut-off between low- and high-affinity clonotypes. The resulting dynamics for $E_a$ then follow the drift-diffusion equation
\begin{equation}
\dot{E}_a(t) = -(g(\phi_a)-\bar{g}(t)) + \sqrt{2D}\eta_a(t),
\end{equation}
from which we observe how $g(\phi_a)$ captures the effect of selection pressure. 

If a clonotype's serum fraction reduces significantly such that $f_a(t)<f_{\min}$, we remove this clonotype from the mixture, effectively terminating its lineage consistent with observations of germinal center selection \cite{DeWitt2025}. To keep the number of clonotypes constant, we then select another clonotype (say $a'$) with probability proportional to $f_{a'}(t)$ and split this to form two new clonotypes with serum fractions $f_{a'}\eta$ and $f_{a'}(1-\eta)$ where $\eta$ is uniformly distributed between 0 and 1. Each new clonotype has affinity $\phi_{a'}$ and these affinities evolve independently at future times. \add{We hold the number of clonotypes fixed and introduce a splitting rule to capture two biological effects that pure replicator dynamics cannot: the effective extinction of low-affinity clonotypes during maturation, and the tendency of large, expanded clonotypes to diversify into sub-populations with distinct affinities. This enables a narrowing of $\PE$ over time.}

We now show how this model can track how $\mathcal{P}(E,t)$ evolves to produce different statistics for $\mathcal{F}$ as observed in Fig.\,\ref{fig:prepost}(c). Taking the green curve from Fig.\,\ref{fig:prepost}(a) as our initial condition (Fig.\,\ref{fig:am}(a), green), we evolve the distribution under the dynamics for $\phi_a$ and $f_a$ and show how the resulting $\PE$ narrows (Fig.\,\ref{fig:am}(a), blue). For the initial and final distributions for $\PE$, we generate distributions for $\mathcal{F}$ (solid lines in Fig.\,\ref{fig:am}(b)) which agree with the experimental data (circular points in Fig.\,\ref{fig:am}(b), same as points in Fig.\,\ref{fig:prepost}(c)): as the distribution $\PE$ gets narrower as a result of affinity maturation, the resulting distribution of $\mathcal{F}$ approaches a Gaussian.

To set the diffusion coefficient $D$ describing somatic hypermutation, we observe that the variance in the change of affinity for clonotype $a$ after $t$ days over many realizations is set by $2Dt$. While recent experimental work \cite{DeWitt2025} quantifying the change in affinities of B-cells during affinity maturation suggests that mutations are common ($\sim0.5/$day on average) and some are sizable (35.7\% lead to a change in affinity of $\Delta\phi_a>0.7$), we choose $D=0.01$\,day$^{-1}$ as our dynamics describe the change in the average affinity within clonotype $a$ across a large number of B-cells trajectories, thus we argue it is significantly less than that of single cells. For $N=1,000$, we set $f_{\min}=1/(10N)=10^{-4}$ and using the parameterization $g(\phi) = -r\tanh((\phi-\phi_c)/\alpha)$ we fit $r=0.2$, $\alpha=1$ and $\phi_c=0.2$ when mapping pre- to post-vaccination $\mathcal{P}(\mathcal{F})$ in Fig.\ref{fig:am}(b).

\end{document}